\documentstyle[12pt]{article}
\title{ Fermionic Path Integrals and Analytic  Solutions for
Two-Dimensional Ising Models}
\author{\large V.\ N.\ Plechko \\[3mm]
\em Bogoliubov Laboratory of Theoretical Physics \\ \em Joint Institute
for Nuclear Research  \\  \em  141980 Dubna, Moscow Region, Russia}

\begin{document} \date{}
\renewcommand{\baselinestretch}{1.0}\pagestyle{plain}
\newcommand{\ba}{\begin{array}}
\newcommand{\ea}{\end{array}}
\newcommand{\eee}{\mbox{e}}
\newcommand{\triex}{\mbox{\hspace{3ex}}}
\newcommand{\edoc}{\end{document}}
\newcommand{\Pfaff}{\mbox{Pfaff}}
\newcommand{\pal}{\partial}
\newcommand{\spsigma}{\ba[t]{c} \mbox{Sp} \vspace{-1ex} \cr
\mbox{$\scriptstyle{(\sigma)}$} \ea }
\newcommand{\spsigmamn}{\ba[t]{c} \mbox{Sp} \vspace{-1ex} \cr
\mbox{$\scriptstyle{(\sigma_{mn})}$} \ea }
\newcommand{\spursigmamn}{\ba[t]{c} \mbox{Sp} \vspace{-1ex} \cr
\mbox{$\scriptstyle{(\sigma_{mn})}$} \ea }
\newcommand{\spura}{\ba[t]{c} \mbox{Sp} \vspace{-1ex} \cr
\mbox{$\scriptstyle{(a)}$} \ea }
\newcommand{\spurasigma}{\ba[t]{c} \mbox{Sp} \vspace{-1ex} \cr
\mbox{$\scriptstyle{(\sigma\,|\,a)}$} \ea }
\newcommand{\spuramn}{\ba[t]{c} \mbox{Sp} \vspace{-1ex} \cr
\mbox{$\scriptstyle{(a_{mn})}$} \ea }
\newcommand{\spurbmn}{\ba[t]{c} \mbox{Sp} \vspace{-1ex} \cr
\mbox{$\scriptstyle{(b_{mn})}$} \ea }

% -----------------------------------------
% --------- start title page  ----------
% -----------------------------------------
\thispagestyle{empty}
\mbox{}\\*[5ex]
HEP-TH/9609044 \\*[5ex]
\begin{center}
{\LARGE  Fermionic Path Integrals and Analytic \\*[2mm]
\mbox{} Solutions for Two-Dimensional Ising Models}
{\large$\dagger$}\vspace*{2ex} \end{center}

\begin{center}
{\large V.\ N.\ Plechko $\ddagger$ \\[2ex]
\em Bogoliubov Laboratory of Theoretical Physics \\[3pt]
    Joint Institute for Nuclear Research \\[7pt]
    141980 Dubna, Moscow Region, Russia} \end{center}

\mbox{} \\ \vfill
\noindent{\large $\dagger$ \ Proceedings of the V International Conference
on  Path Integrals from meV to  MeV, JINR, Dubna, Russia, May 27-31, 1996.
Path Integrals: Dubna$\,$'$\,$96. Edited by  V.S. Yarunin and M.A.
Smondyrev (Publ JINR E-96-321, Dubna, 1996) p.p. 295--299. } \\*[1ex]

\noindent{\large$\ddagger$} \ E-mails:\ plechko@thsun1.jinr.dubna.su,
plechko@theor.jinrc.dubna.su

\vspace*{15ex}

% -----------------------------------------
% --------- start basic text   ---------
% -----------------------------------------
\newpage
\setcounter{page}{295}

\maketitle
\vspace*{-24pt}
\begin{abstract}
The notion of the integral over anticommuting Grassmann variables
(non-quantum fermionic fields) seems to be the most powerful tool in order
to extract the exact solutions for the 2D Ising models on simple and more
complicated lattices, which is the subject of a discussion in this report.
\end{abstract}

{\bf 1. Introduction.}\
The fermionic structures in the two-dimensional (2D) Ising model [1] were
first recognized within the transfer-matrix and combinatorial approaches
[2]. It was realized later on that the notion of the integral over
anticommuting Grassmann variables due to Berezin [3] is a powerful tool to
study the 2D Ising model [3-8]. A simple method of fermionic analysis for
the 2D Ising models disposed on simple and more complicated lattices has
been developed in [7-8]. The approach is based on the integration over the
anticommuting Grassmann variables (nonquantum fermionic fields) and the
mirror-ordering factorization principle for the 2DIM density matrix. The
method is straightforward, the traditional transfer-matrix or combinatorial
considerations are not used.  Schematically, we have:  $$  \ba{llr}
Q\,=\,\spsigma\,Q\,(\sigma) \to\spurasigma\,Q\, (\sigma\,|\,a)\to
\spura\,Q\,(a)\,=\,Q\,.  \ea\eqno(1) $$ We start here with the original
Ising spin partition function, $Q$, and introduce, in a special way, a set
of new purely anticommuting Grassmann variables $(a)$ thus passing to a
mixed $(\sigma\,|\,a)$ representation.  Eliminating spin variables
$(\sigma)$ in this mixed representation, we obtain a purely fermionic
expression for the same partition function $Q$.  The final expression for
$Q$ appears as a fermionic Gaussian integral. In essence, this means the
exact solution of the problem. The partition function is expressible here
as a fermionic Gaussian integral even for the most general inhomogeneous
distribution of coupling parameters [7]. In particular, for the standard
rectangular lattice this gives a few line derivation of the Onsager result
[7]. The 2D Ising lattices with more complicated local structures have been
analyzed by the factorization method as well [8,9]. The Gaussian fermionic
representations have recently been constructed also for the inhomogeneous
2D dimer models [10]. In this scheme, we introduce Grassmann variables in
order to decouple the bond Boltzmann weights into separable factors called
the Grassmann factors (GFs). We then combine GFs with the same spin
variables into separable groups and sum over spins in each group
independently thus passing to a purely fermionic representation for $Q$.
The key point of all the construction is the mirror-ordering procedure for
the global products of GFs which enables one to perform the actual summing
over spin degrees of freedom [7,8]. In what follows we comment shortly on
some aspects of the method.\\*[-2ex]

%
%           2. grassmann variables
%
{\bf 2. Grassmann variables.}\ Grassmann variables are the purely
anticommuting fermionic numbers. Given a set of Grassmann variables
$a_1,a_2, ..., a_N$, we have $a_ia_j+a_ja_i=0$, $a_{j}^{2}=0$. Berezin's
rules of integration for one variable are [3]:
$$
\ba{llr}
\int da_j \cdot a_j = 1\,, \triex\triex
\int da_j \cdot 1   = 0\,. \triex\,
\cr\ea\eqno(2)
$$
In the multidimensional integral, the differentials $da_1,da_2,...,da_N$
are again anticommuting with each other and with the variables. The basic
relations of the grassmannian analysis concern the Gaussian fermionic
integrals [3]. The Gaussian integral of the first kind is related to
the determinant:
$$
\ba{llr}
\int\,\prod\limits_{j=1}^{N}\,da_{j}^{\,*}da_{j}\,\exp\left(
\sum\limits_{i=1}^{N}\sum\limits_{j=1}^{N}a_{i}A_{ij}a_{j}^{\,*}
\right)\,=\,
\det\hat{A}\;,
\cr\ea\eqno(3)
$$
where $\{a_{j},a_{j}^{\,*}\}$ is a set of completely anticommuting
Grassmann variables, the matrix in the exponential is arbitrary. The
fermionic exponential here is assumed in the sense of its series expansion,
the series terminates at some stage due to the property $a_{j}^{\,2}=0$.
The origin of the determinant in (3) is due to the known interrelation
between the fermionic algebra and determinant combinatorics. By convention,
the variables $a_{j}^{}$ and $a_{j}^{\,*}$ can be considered as complex
conjugated fermionic fields, otherwise these are simply independent
variables. The Gaussian integral of the second kind, for real fermionic
fields, is related to the Pfaffian:
$$
\ba{llr} \int\,da_N\,...\,da_2da_1\,\exp\left(
\frac{1}{2}\,\sum\limits_{i=1}^{N}\sum\limits_{j=1}^{N}
a_{i}A_{ij}a_{j}\right)\,=\,\mbox{Pfaff}\,\hat{A}\,,
\hspace{3ex} A_{ij}=-A_{ji}\,,
\cr\ea\eqno(4)
$$
where the matrix is assumed to be skew-symmetric. The Pfaffian is some
combinatorial polynomial in elements $A_{ij}^{}$ known in mathematics for
a long time. Pfaffian combinatorics is in fact identical with that of the
fermionic version of Wick's theorem. For any skew-symmetric matrix $(A_{ij}
= - A_{ji})$ we have: $\;\det \hat{A} = (\,\mbox{Pfaff}\,
\hat{A}\,)^{\,2}\;$.\\*[-1ex]

%
%     3. ising rectangular
%
{\bf 3. The 2D Ising model on rectangular lattice.}\ To illustrate (1),
let us consider the 2D Ising model on a rectangular lattice net with
the inhomogeneous distribution of coupling parameters [7]. The hamiltonian
is given as follows:
$$
\ba{llr}
-\beta\,H\,(\sigma)=\sum\limits_{m=1}^{L}\sum\limits_{n=1}^{L}\;
[\;b_{m+1n}^{(1)}\sigma_{mn}\sigma_{m+1n}+ b_{mn+1}^{(2)}
\sigma_{mn}\sigma_{mn+1}\,]\,, \;\;\beta=1/kT\,,
\cr\ea\eqno(5)
$$
where  $b_{mn}^{(\alpha)} = J_{mn} ^{(\alpha)}/ kT$, $J_{mn}^{(\alpha)}$
are the magnetic exchange energies, $\beta=1/kT$ is the inverse
temperature.  The Ising spins $\sigma_{mn}= \pm1$ are disposed at lattice
sites $mn$, with $m,n=1,...,L$, $N=L^{\,2} \to \infty$. For finite $N$, we
assume free-boundary conditions:  $\sigma_{M+1n}=\sigma_{mN+1} =0$. Noting
that for typical bond weight $\eee^{\,b\sigma \sigma'}= \cosh b + \sinh b\,
\sigma \sigma'$, which readily follows from $(\sigma \sigma')^{2} = 1$, we
come to the reduced partition function:
$$
\ba{llr}
Q =\spsigma\{\,\prod\limits_{mn}\,(\,1+t_{m+1n}^{(1)}\sigma_{mn}
\sigma_{m+1n})\,(1+t_{mn+1}^{(2)}\sigma_{mn}\sigma_{mn+1})\,\}\,,\triex
\ea\eqno(6)
$$
where $t_{mn}^{(\alpha)} = \tanh\,b_{mn}^{(\alpha)}$, and $Sp_{(\sigma)} =
2^{\,-N} \Sigma_{ (\sigma)}$ stands for a properly normalized spin
averaging [7]. Following (1), we introduce a set of Grassmann variables
$\{a_{mn}$, $a_{mn}^{\,*}$, $b_{mn}$, $b_{mn}^{\,*}\}$, a pair per bond,
and factorize the local bond weights as follows:
$$
\ba{llr}
1+t_{m+1n}^{(1)}\sigma_{mn}\sigma_{m+1n}
=\int\limits_{}^{}da_{mn}^{*}da_{mn}\,\mbox{e}^{\,a_{mn}a_{mn}^{*}}\,
(1+a_{mn}^{}\sigma_{mn}^{})\,
(1+t_{m+1n}^{(1)}\,a_{mn}^{*}\sigma_{m+1n}^{})\,,
\cr  1+t_{mn+1}^{(2)}\sigma_{mn}\sigma_{mn+1}
=\int\limits_{}^{}db_{mn}^{*}db_{mn}\,\mbox{e}^{\,b_{mn}b_{mn}^{*}}\,
(1+b_{mn}^{}\sigma_{mn}^{})\,
(1+t_{mn+1}^{(2)}\,b_{mn}^{*}\sigma_{mn+1}^{})\,.
% \vspace*{-3ex}\cr
\ea\eqno(7)
$$
These identities can be checked simply by using the rules (2), and noting
that $\eee^{\,aa^{\,*}} = 1 + aa^{\,*}$, since $(aa^{\,*}) ^{\,2} =0$.
Neglecting the sign of the Gaussian fermionic averaging, we see that the
bond weights are presented now as $A_{mn}^{}A_{m+1n}^{\,*}\,$, $\,B_{mn}^{}
B_{mn+1} ^{\,*}$, where the Grass\-mann factors $\,A_{mn}^{}\,,\,
A_{m+1n}^{\,*}\,,\, B_{mn}^{}\,,\, B_{mn+1}^{\,*}\,$ are to be identified
>from (7).

At the next stage, we group together, over the whole lattice, the four
factors with the same Ising spin $\sigma_{mn}$. These four factors come by
the factorization of the four different bonds (weights) attached to a given
$mn$ site. Performing the averaging over $\sigma_{mn} \pm1$ in each group,
we find [7]:
$$
\ba{llr} \spursigmamn\!\!\{\,A_{mn}^{\,*}B_{mn}^{\,*}A_{mn}B_{mn}\,\} =
\frac{1}{2}\!\!\sum\limits_{\sigma_{mn}=\pm1}^{}\!\!\!
(1+t_{mn}^{(1)}\,\sigma_{mn}a_{m-1n}^{\,*})\,
(1+t_{mn}^{(2)}\,\sigma_{mn}b_{mn-1}^{\,*})\,
\cr
\times (1+\sigma_{mn}a_{mn}^{})\,(1+\sigma_{mn}b_{mn}^{}) =
1+t_{mn}^{(1)}t_{mn}^{(2)}\,a_{m-1n}^{\,*}b_{mn-1}^{\,*}+a_{mn}b_{mn}+
\cr
+\, (t_{mn}^{(1)}a_{m-1n}^{\,*} + t_{mn}^{(2)}b_{mn-1}^{\,*})\,
(a_{mn}+ b_{mn})\,+\,
t_{mn}^{(1)}t_{mn}^{(2)}\,a_{m-1n}^{\,*}b_{mn-1}^{\,*}a_{mn}b_{mn} =
\cr
=\exp\,[a_{mn}b_{mn}+t_{mn}^{(1)}t_{mn}^{(2)}a_{m-1n}^{\,*}
b_{mn-1}^{\,*}+(t_{mn}^{(1)}a_{m-1n}^{\,*} + t_{mn}^{(2)}b_{mn-1}^{\,*})\,
(a_{mn}+ b_{mn})]\,.
\ea\eqno(8)
$$
Taking also into account the diagonal Gaussian terms arising under
factorization (7), we obtain the partition function (6) in the form [7]:
$$
\ba{llr}
Q\;=\; \int \prod\limits_{m=1}^{L}\prod\limits_{n=1}^{L}
db_{mn}^{\,*}db_{mn}^{}da_{mn}^{\,*}da_{mn}^{}\;
\exp\,\;\{\;\sum\limits_{m=1}^{L}\sum\limits_{n=1}^{L}\,[\,
a_{mn}^{}a_{mn}^{\,*}+b_{mn}^{}b_{mn}^{\,*}+
\cr +\,
a_{mn}^{}b_{mn}^{}+t_{mn}^{(1)}t_{mn}^{(2)}a_{m-1n}^{\,*}b_{mn-1}^{\,*}+\,
(t_{mn}^{(1)}a_{m-1n}^{\,*}+t_{mn}^{(2)}b_{mn-1}^{\,*})\,(a_{mn}^{}+
b_{mn}^{})\,]\,\}\;,
\ea\eqno(9)
$$\\
with the free-boundary condition for fermions: $a_{0n}^{\,*}=0$,
$b_{m0}^{\,*} =0$. Now $Q$ is given as a Gaussian fermionic integral,
equivalently, the 2D Ising model is reformulated as a free fermion field
theory on a lattice. For space restriction, we do not comment here on the
mirror-ordering procedure which just makes it possible to put the four
relevant factors nearby as in (8), see for details [7,8]. It is
important that the method works for the inhomogeneous lattices, this might
be of interest, in particular, for studies of random systems [6,11]. The
integral (9) can in fact be simplified integrating out the
$a_{mn}^{},b_{mn}^{}$ fields by means of the identity $\int db\,da\,
\exp\,(ab + aL + L^{\,'}b) = \exp\,(LL^{\,'})$, where $L,L^{\,'}$ are some
linear forms in Grassmann variables independent of $a,b$.\\*[-2ex]

%
%     4. complicated lattices
%
{\bf 4. The 2D Ising models on complicated lattices.}\ An important
modification of the method has been introduced in [8], where we start with
the factorization of the {\em cell}\ weights presented by three-spin
polynomials. The spin polynomial interpretation arises if we multiply few
local weights forming elementary cell in $Q$. This enables us to obtain the
fermionic representation for $Q$ with only two fermionic variables {\em per
site}, which provides essential simplifications in the analysis [8]. For a
set of 2D Ising models, including the standard rectangular, triangular and
hexagonal lattices as the simplest cases, the partition function then
reduces to the following expression [8]:
$$
\ba{llr}
Q=\spsigma \{\,\prod\limits_{mn}^{}\,(\alpha_0 + \alpha_1\,
\sigma_{mn}\sigma_{m+1n}+ \alpha_2\,\sigma_{m+1n}\sigma_{m+1n+1}+
\alpha_3\,\sigma_{mn}\sigma_{m+1\,n+1})\,\}\,,
\cr\ea\eqno(10)
$$
where $\alpha_0$, $\alpha_1$, $\alpha_2$, $\alpha_3$ are numerical
parameters specific for each lattice (we now assume the homogeneous case).
The three-spin polynomial in (10) is the effective Boltzmann weight of the
elementary cell. The factorization method then yields the following
Gaussian fermionic integral for the partition function [8]:
$$
\ba{lll}
Q\, = \int \prod\limits_{mn}^{}dc_{mn}^{\,*}dc_{mn}^{}\exp\,\{\,
\sum\limits_{mn}^{}\,[\,\alpha_{0}\,c_{mn}c_{mn}^{\,*} -
\alpha_1\,c_{mn}c_{m-1n}^{\,*} - \alpha_2\,c_{mn}c_{mn-1}^{\,*}-\cr
\hspace*{12ex}
-\,\alpha_{3}\,c_{mn}c_{m-1n-1}^{\,*} -\alpha_{1}\,c_{mn}c_{m-1n}-\,
\alpha_{2}\,c_{mn}^{\,*}c_{mn-1}^{\,*}\,]\,\}\;,
\cr\ea\eqno(11)
$$
where $c_{mn}^{}, c_{mn}^{\,*}\,$ is a set of the purely anticommuting
Grassmann variables (two per cell). The derivation of (11) is simple [8].

The explicit evaluation of the integral (11) can be performed by passing to
the momentum space (Fourier substitution for fermions). Taking then the
limit of infinite lattice, we find the free energy per cell as follows
[8]:
$$
\ba{llr}
-\beta f_Q = \,\lim_{\,L\rightarrow\infty}^{}\,
(\frac{1}{L^2}\ln Q\,)\,
=\cr =  \frac{1}{2}\,\int\limits_{0}^{2\pi}\int\limits_{0}^{2\pi}
\frac{dp}{2\pi} \frac{dq}{2\pi}\,\ln\,[\,
(\alpha_{0}^{2}+\alpha_{1}^{2}+\alpha_{2}^{2}+\alpha_{3}^{2}) -
2\,(\alpha_{0}^{}\alpha_{1}^{}-\alpha_{2}^{}\alpha_{3}^{})\,\cos p\,-
\cr -\,
2\,(\alpha_{0}^{}\alpha_{2}^{}-\alpha_{1}^{}\alpha_{3}^{})\,\cos q\,-
2\,(\alpha_{0}^{}\alpha_{3}^{}-\alpha_{1}^{}\alpha_{2}^{})\,\cos\,
(p+q)\,]\;.
\cr\ea\eqno(12)
$$
The symmetries of this solution, and the closely related question on the
location of the critical point, have an interesting interpretation within
the spin-polynomial approach, as is discussed in [8], also see [12]. By
a suitable specification of the parameters $\alpha_0,\alpha_1,\alpha_2,
\alpha_3$ from (12) we obtain, in particular, the solutions for the free
energies for canonical rectangular, triangular, and hexagonal Ising
lattices, as well as for some other lattices with more complicated local
structures [8]. For application of these results to the analysis of the
regularly diluted 2D Ising ferromagnets also see [9]. The fermionic
integral for $Q$ with the minimal number of the fermionic components (11)
appears to be a suitable starting point to formulate the continuum-limit
field theories for 2D Ising models near $T_c$, as is discussed in more
detail in [12]. The resulting field theory is the massive two-component
Majorana theory in the 2D euclidean space-time [12]. By doubling of
fermions, we can pass as well to the 2D Dirac field theory of charged
fermions [12].\\[-2ex]

{\bf 5. Conclusions}.\ We have discussed some aspects of a simple
non-combinatorial approach to the analytic solutions for the 2D Ising
models on simple and more complicated lattices. For any planar 2D Ising
model the partition function can be expressed as a fermionic Gaussian
integral. The integral over the anticommuting Grassmann fields is a
powerful tool to analyze the two-dimensional Ising models.
% \vspace*{-2ex}

\end{document}